\documentclass{ws-p8-50x6-00}

\begin{document}

\title{QCD Processes at the LHC}

\author{Georgios D. Stavropoulos}

\address{CERN}

\maketitle

\abstracts{
The study of QCD processes at the LHC will serve two main goals. First the
predictions of QCD will be tested and precision measurements will be
performed. Second QCD processes represent a major part of the background to
other Standard Model processes and signals of new physics at the LHC and thus
need to be understood precisely in the new kinematic region available here.
Furthermore, the production cross-sections for almost all processes are
controlled by QCD.
}

\section{Introduction}
The Large Hadron Collider (LHC) is a proton-proton collider with 14 TeV center
of mass energy and design luminosity of 10$^{34}$ cm$^{-2}$s$^{-1}$. Beam
crossings are 25 ns apart and at design luminosity there are on average 23
interactions per crossing. 10 fb$^{-1}$ of integrated luminosity are expected
to be collected in one year of data taking at the initial low (10$^{33}$
cm$^{-2}$s$^{-1}$) luminosity and 100 fb$^{-1}$ for one year of data taking at
the nominal one.
 
A detailed understanding of QCD is important for almost all the physics
processes to be studied at the LHC, as the production mechanisms are mostly
controlled by QCD. LHC will allow QCD studies to be performed at very high
energy, including precision tests and measurements in an as yet unexplored
kinematic region. 

\begin{figure}[t]
\epsfxsize=12pc 
\epsfbox{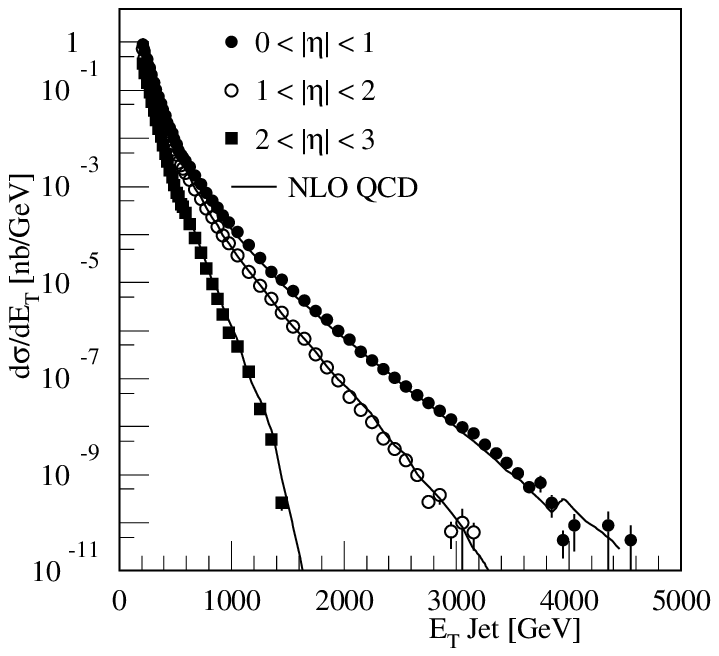}
\epsfxsize=12pc
\epsfbox{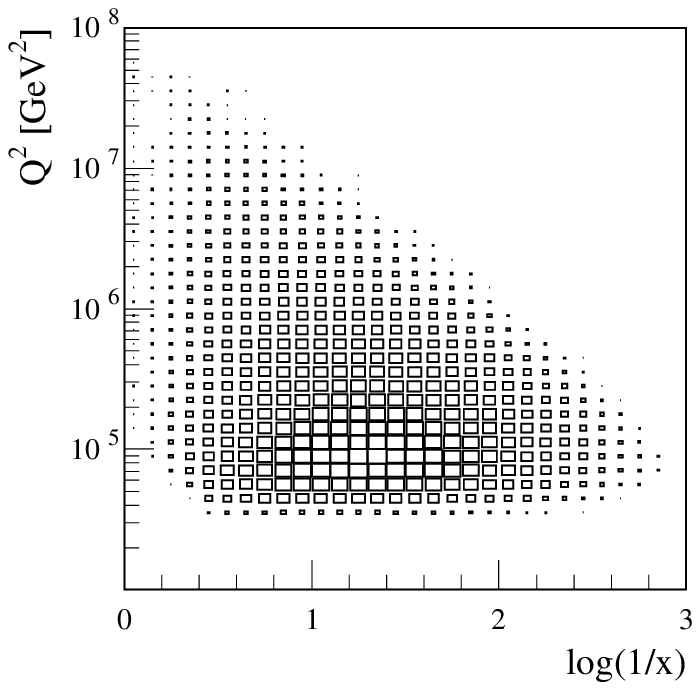}
\caption{{\em (a):} Inclusive jet cross-section as a function of the transverse
  energy of the jet.  {\em (b):} Range in $1/x$ and $Q^2$ for the di-jet
  differential cross-section measurement. \label{fig:jets}}
\end{figure}

\section{Jet Physics}
The measurement of jet production cross-section at LHC will provide a
stringent test of perturbative QCD in an energy regime never probed so far. 
In Figure \ref{fig:jets}{\em (a)} the inclusive jet cross-section\cite{at} is
shown as a function of the transverse energy of the jet, $E_T$, for three
different bins in $\eta$. The expected statistics for an integrated luminosity
of 30 fb$^{-1}$ amounts to 4$\cdot$10$^5$ events with $E_T^{jet} >$ 1 TeV,
demonstrating the fact that at LHC the statistical uncertainties on the jet
cross-section measurement will be small. More important in this measurement
are the systematic errors, which can arise from the jet algorithm, the
knowledge of the jet energy scale and the understanding of the overall
calorimeter response, the knowledge of the jet trigger efficiency, the
knowledge of the luminosity for the overall normalization, and the effect of
the underlying event. The expected statistics at large jet $E_T$ values
implies a desirable control of the systematic uncertainties to a precision of
better than 1\% for $E_T <$ 1 TeV and to about 10\% for $E_T$ of about 3 TeV.
The jet cross-section measurement can be used for the determination of
$\alpha_s$. It was shown\cite{as} that at LHC a measurement of $\alpha_s(M_Z)$
up to scales of order TeV can be achieved with a 10\% accuracy.

The measurement of di-jet events and their properties can be used to constrain
the parton densities inside the proton, in various kinematic regions of
$Q^2$ and $x$. In Figure \ref{fig:jets}{\em (b)} the expected range in $x$ and
$Q^2$ accesible with the di-jet differential cross-section measurement is
shown. For a transverse energy threshold of 180 GeV , most of the events have
$Q^2$ values of about 10$^5$ and values of $x$ between 2$\cdot$10$^{-3}$ and
0.5.

\begin{figure}[t]
\epsfxsize=12pc 
\epsfbox{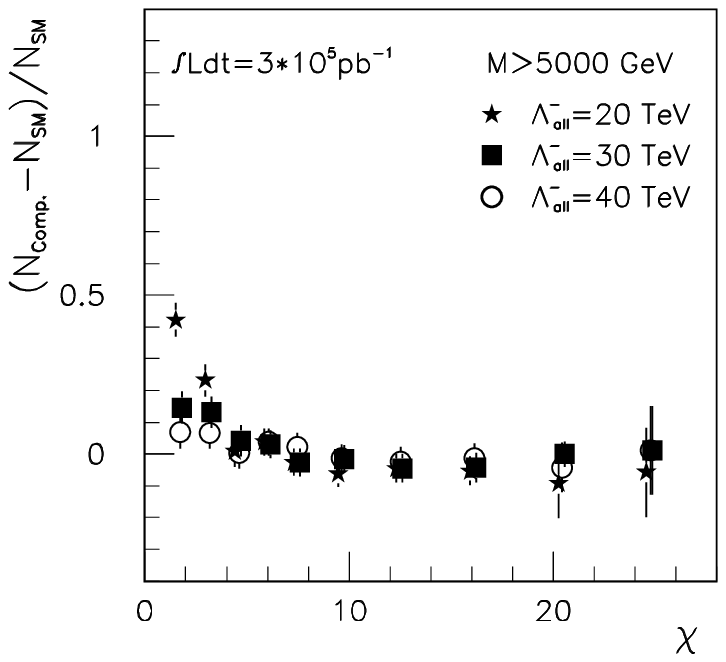} 
\epsfxsize=12pc
\epsfbox{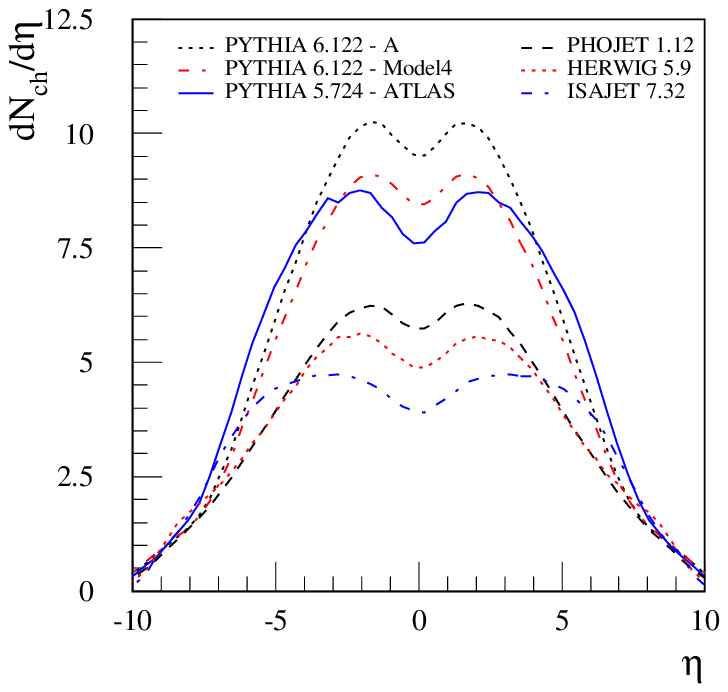}
\caption{{\em (a):} Deviation of the di-jet angular distribution from the
  Standard Model predictions, for di-jet mass above 5 TeV for 300 fb$^{-1}$.
  {\em (b):} Charged particle density in minimum-bias events at LHC energies
  as a function of pseudorapidity, for six model predictions. \label{fig:cmb}}
\end{figure}

The di-jet invariant mass and angular distributions can be used to search for
new physics. Studies have shown that the di-jet angular distribution has an
excellent discovery capability for quark compositeness. An integral luminosity
of 300 fb$^{-1}$ will allow excluding quark substructure at 95\% CL, if
the constituent interaction constant is of the order of 40 TeV (see figure
\ref{fig:cmb}{\em (a)}).

\section{Photon Physics}
Direct photon measurements can provide important constraints on parton
distributions, especially on the gluon distribution in the proton. 
The detection of photons at hadron collider is a challenging task due to the
large background from jet production, where fluctuations can mimic the
signature of a photon. As an example in ATLAS it was found that for a photon
efficiency of 80\% a rejection factor against jets of larger than 3x10$^3$ can
be achieved for $E_T >$ 40 GeV, thus providing a good signal-to-background
ratio. The selection at ATLAS will be based on a trigger demanding an isolated
photon of $E_T >$ 60 GeV within $|\eta| <$ 2.5. This gives a value of the
Bjorken-$x$ of 5$\cdot$10$^{-4}$.

\section{Lepton Physics}
The measurement of the Drell-Yan lepton pair production and the production of
$W$ and $Z$ bosons (with a leptonic decay to electrons or muons) will allow
one to constrain the quark and anti-quark densities of the proton at a scale
given by the invariant mass of the lepton pair ($W/Z$ mass respectively) over
a wide range in Bjorken-$x$. For example, the measurement of $Z$ boson
production cross-section allows one to cover the range of
3$\cdot$10$^{-4} < x <$ 0.1 in Bjorken-$x$ at $Q^2$ = 8$\cdot$10$^3$ GeV$^2$.

\section{Hard Diffractive Scattering}
During the last few years the understanding of diffractive phenomena has
received revived attention due to the observation of hard diffractive events
in several experiments. Thus, the study of hard diffractive scattering at LHC
becomes an important topic of the physics program of the involved experiments.
An extension of the LHC detectors in the forward region (beyond
pseudo-rapidities of $|\eta|$ = 5) is presently under study, in order to
increase the acceptance for charged particles from inelastic interactions
(possibly including a momentum measurement) and to provide tagging and
measurement of leading protons from elastic and diffractive interactions.

\section{Properties of minimum-bias events}
At the LHC, minimum-bias events will make up the 23 interactions per bunch
crossing at high luminosity. In order to understand precisely their
contribution to the measured quantities for the hard scattering events of
interest, a detailed knowledge of the structure of the minimum-bias events is
required. In Figure \ref{fig:cmb}{\em (b)} the charged particle density in
minimum-bias events at LHC energies is shown as a function of pseudorapidity.
Six models, tuned on CDF data, were used for this calculation. There is a
difference of a factor 2 in the expected charged particle density in the
central region, for models which are able to describe CDF data. Each
minimum-bias event is expected to contribute from charged particles about
0.5 to 1 GeV per unit rapidity and unit azimuth to the transverse energy.

\section{Conclusions}
The LHC will provide a large sample of events with high $p_T$ signatures for
QCD studies. It will extend the kinematic range to values of $Q^2$, the
hard scale of the partonic process, of the order of TeV$^2$.
The fraction of the proton momentum attributed to a parton will reach values
below 10$^{-5}$, while keeping the hard scattering scale above 100 GeV$^2$. 

A precise knowledge and understanding of QCD processes is important for the
studies of the Higgs boson(s) and for searches for new physics beyond the
Standard Model, where QCD represents a large part of the background.

\end{document}